# RADIOACTIVE ION BEAM FACILITIES IN EUROPE: CURRENT STATUS AND FUTURE DEVELOPMENT

John C. Cornell, GANIL, BP 55027, Caen 14076 cedex 5, France

*Abstract*

The production and acceleration of Radioactive Ion Beams (RIBs) is today an area of intense interest. The history and development of RIB facilities in Europe is presented, with discussion of both the "in-flight" and "ISOL" methods used at different laboratories. The current status of present developments like the SPIRAL II facility at GANIL and the FAIR in-flight facility at GSI are briefly reviewed. Particular emphasis is given to the recent EURISOL Feasibility Study for a European ISOL facility, and the present EURISOL Design Study.

## INTRODUCTION

A cynic once said to me, "Ah, they've done all the interesting nuclear physics with stable beams, so now they have to use radioactive ion beams!" While for some this may hold a grain of truth, there is far more to it than that. The quest for greater understanding of the properties of nuclei, and especially the way they were formed in the early Universe, leads us to explore the outer regions of the chart of the nuclei. This implies capturing exotic nuclei (in traps), especially those close to the so-called proton and neutron "drip-lines" and, where possible, creating enough of these nuclei to accelerate beams of them for studying their interactions with target nuclei.

The problem, of course, is that radioactive nuclei are unstable, and the more exotic they are, the shorter their life-times. The battle which faces the accelerator physicist is thus the need for high enough yields of such unstable nuclei despite their generally low production cross sections and extremely short half-lives. While accelerated particle beams with enough energy will easily produce enormous quantities of radioactive nuclei, the difficulty is to analyse these and select only the desired nuclei, often orders of magnitude less numerous than the more common nuclei that are produced simultaneously.

## SOME EARLY HISTORY

The history of radioactive ion beams (RIBs) [1] is now more that 50 years old [2]. Otto Kofoed-Hansen and Karl Ove Nielsen at the Niels Bohr Institute in Copenhagen were the first to perform such an experiment [3] in 1951. They were investigating beta-decay and neutrino emission from neutron-rich krypton isotopes, produced in uranium fission. Using a variant of the "converter" method – now enjoying renewed interest for the next generation of RIB facilities – they used deuterons from their cyclotron on an internal target to produce neutrons. These then struck a uranium oxide target (mixed with baking powder) from which gas flowed. The gas was then ionised, extracted from a high-voltage platform and passed through a mass-separator, which selected the krypton ions of interest.

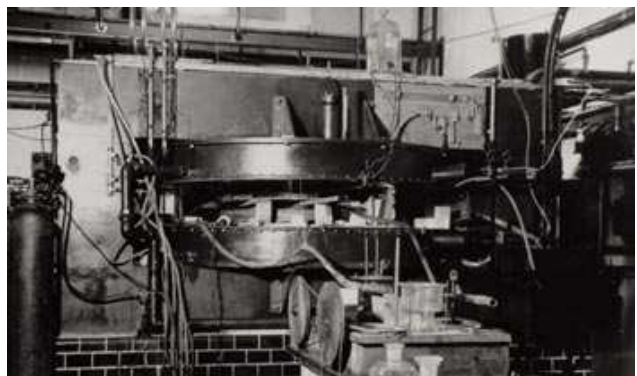

Figure 1: The Niels Bohr Institute cyclotron.

This cyclotron (Fig. 1) was later moved from the Niels Bohr Institute, ending these experiments, but Nielsen and Kofoed-Hansen were also active at CERN, and in 1967 the ISOL method was introduced at ISOLDE, using proton beams accelerated in the Synchro-Cyclotron (SC)

In 1973, ISOLDE was reconstruction to accommodate the increased intensity of the 4-µA, 600-MeV external proton beam from the SC. When the SC finally closed down in 1990, a new ISOLDE site was already being constructed at CERN, to be fed with up to 2 µA of 1-GeV protons from the Proton Synchrotron Booster (PSB), and the first experiments took place in 1992. [4] In 1999 the facility was upgraded to handle 4-µA proton beams at 1.4 GeV. We shall return to ISOLDE in a later section.

Many other laboratories have since made use of this method of producing and analysing RIBs, with extracted beam energies around 60 keV (derived from a high-voltage platform), suitable for use with high-resolution mass spectrometers. The exploitation of various ion-source techniques has been continuously reported in the EMIS series of conferences in recent years [5-9].

## ISOL AND IN-FLIGHT METHODS

### *"Classical" ISOL facilities*

The ISOL ("ion-source-on-line") method requires a high-intensity primary beam of light particles from a "driver" accelerator (or a reactor), and a thick hot target, from which the exotic nuclei formed have to diffuse and effuse into an ion source, for ionisation and extraction (Fig. 2). As this method generally results in the formation of a whole host of nuclei, stable and non-stable, it is essential to use mass separators to select specific ions of interest. Since very high intensities of driver beams of high-energy protons (or reactor neutrons) are available, exotic nuclei with extremely low production cross sections can still be obtained in observable numbers. However, some isotopes have half-lives which are just too short for enough nuclei to survive the ISOL method.

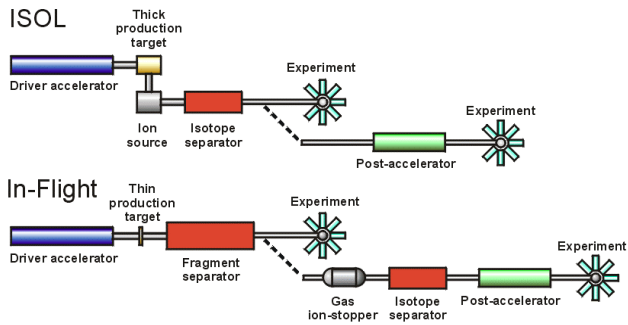

Figure 2: Comparison of ISOL and "in-flight" methods of RIB creation and post-acceleration.

Many facilities now exist for RIB production using the ISOL technique: in Europe alone there are "classical" facilities at CERN, Jyväskylä, Studsvik, Louvain-la-Neuve, Orsay, Warsaw, Dubna and elsewhere. There are excellent review articles on the ISOL method, including that by Ravn [10] and more recently by Lindroos [11].

The beams of radioactive ions resulting from the ISOL method have excellent optical properties, since these depends only on the ion source, and such high-quality beams are excellent for mass and isobar selection in mass spectrometers without unusually high acceptance.

However, the dependence on the chemistry of the elements produced makes it hard to extract some elements from the target, or to ionise sufficient numbers, such as the notoriously difficult refractory elements, for example.

*The "In-flight" method*

The "in-flight" method uses fragmentation of intense heavy-ion beams, generally in a thin target, in which the forward momentum of the primary beam fragments is exploited for mass separation and study or further reactions. These beams are independent of chemistry, but as a result of the interaction in the target, they do not have good optical qualities, and need carefully designed mass separators with high acceptance. Since the intensities of the heavy-ion beams are generally lower than that of the light-ion beams used for the ISOL method, the yields of some exotic fragments may also be somewhat lower.

In-flight facilities – sometimes called fragmentation facilities – have been reviewed by many authors, including Morrissey and Sherrill [12].

The in-flight technique was first developed in the 1970s at the Bevalac accelerator at the Lawrence Berkeley Laboratory (LBL, now LBNL) [13] in the USA.

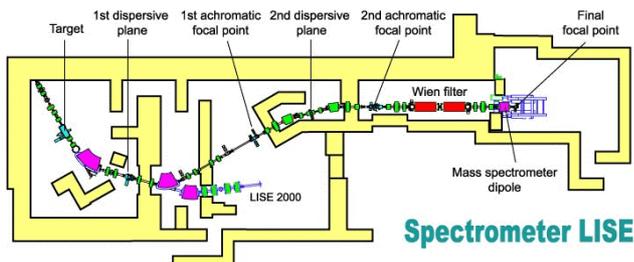

Figure 3: The LISE spectrometer at GANIL.

Later it was greatly improved at the LISE separator at GANIL with the introduction of degraders [14]. This separator (Fig. 3) uses a symmetrical 2-dipole magnetic analyser which selects in $A \times V/Z$, and the energy-loss in a degrader between the dipoles selects in $A^3/Z^2$, after which a Wien filter selects for the velocity $V$, and a final dipole selects in $A/Z$. The target in the LISE beamline receives beams of heavy ions with energies up to 95 A MeV from the chain of successive cyclotrons at GANIL (see Fig. 7).

Other in-flight RIB facilities are SISSI at GANIL, the ETNA fragment separator at LNS in Catania, COMBAS and the ACCULINNA separator at JINR in Dubna.

It is clear that the selection of isobars is essential for unambiguous identification of RIBs. The "FRS" fragment separator at GSI, in Germany (Fig. 5) is a good example of the techniques used with in-flight production of RIBs.

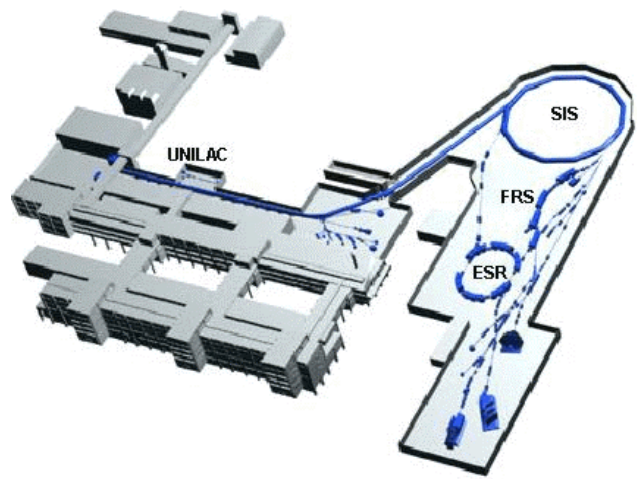

Figure 4: The present layout of the GSI facility.

## ISOL WITH POST-ACCELERATORS

*Louvain-la Neuve*

While the first beams from ISOL technique were simply accelerated from a high-voltage platform, a very different approach began at Louvain-la-Neuve in 1987. In a collaboration between three Belgian universities (UC Louvain, UL Bruxelles and KU Leuven), this project aimed at producing intense, energetic radioactive ion beams, using a cyclotron as a post-accelerator. After some extensive plumbing to lead the radioactive ions produced from the ion source to the cyclotron, the first $^{13}$N beam was accelerated in 1989.

The first cyclotron, CYCLONE 30, is a 30-MeV proton accelerator and is used to produce the desired exotic element in suitable targets [15]. Today, beam intensities up to 300 μA (depending on target and desired charge state) are used to irradiate the production target.

This resulting activity diffuses out of the hot target and is fed into a 6.4-GHz ECR ion source, designed to produce low charge states with a high efficiency. After a first magnetic separation in a low mass resolution dipole, the ions are injected in the second cyclotron, CYCLONE 110, for subsequent acceleration up to 0.65–5 MeV/u.

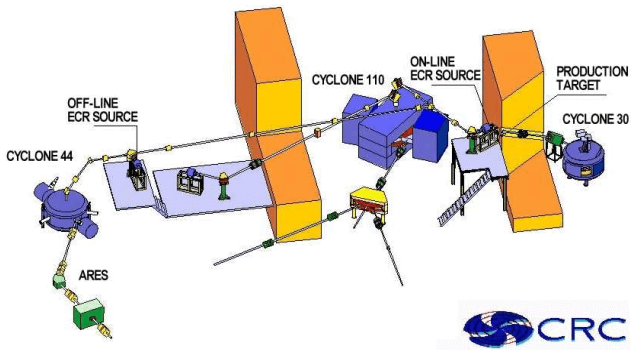

Figure 5: The Louvain-la-Neuve facility today

To remove any isobaric contamination from the beam, the cyclotron is tuned as a radiofrequency mass spectrometer, simply by reducing the acceleration voltage which results in an increased number of turns in the machine. This allows separation of beams with a relative mass difference of $2\times10^{-4}$ (e.g. enough to separate $^{19}$Ne from $^{19}$F). This method results in an isobaric suppression factor of $10^{-6}$ while maintaining a transmission through the cyclotron (acceleration and separation) of 3 to 5%.

A third cyclotron (CYCLONE 44) is used for increased transmission efficiency of the low-energy ISOL beams (below 1 MeV/nucleon) needed for nuclear astrophysics.

## CERN-ISOLDE

With the high-energy, high-current proton beams (up to 4 µA) available from the PS Booster, researchers at ISOLDE have produced many new exotic isotopes, and have led the world in RIB production. Today, more than 700 different isotopes can be produced, from 25 different target materials, using 4 families of ion sources.

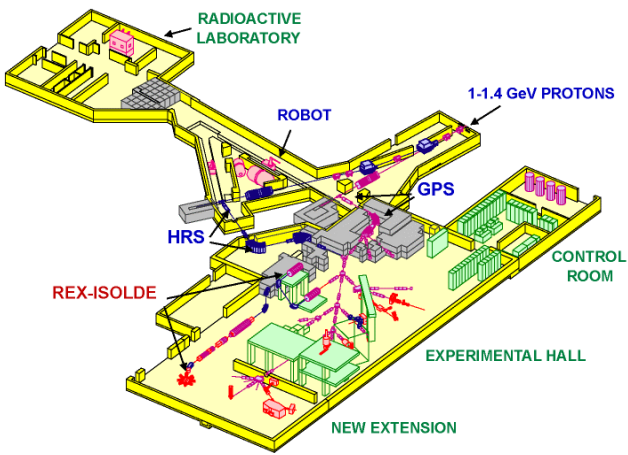

Figure 6: The ISOLDE facility at CERN

At CERN, a linear accelerator REX-ISOLDE [16] is used to post-accelerate the RIBs produced at ISOLDE. The exotic nuclei are first singly-ionised and separated in the on-line mass separator, accumulated in a Penning trap, then charge-bred in an electron-beam ion source to enhance the charge state of the radioactive ions to $A/q \leq 4.5$. Finally these ions are accelerated in the linac, originally to 2.2 MeV/u, but recently upgraded to 3.1 MeV/u. There are plans to increase this to ~5 MeV in the near future.

## SPIRAL at GANIL

SPIRAL is an ISOL facility at GANIL, with heavy-ions from the coupled cyclotrons fed into a RIB production facility [17], after which the exotic nuclei are injected into the CIME cyclotron and post-accelerated, before being directed to existing user areas (Fig. 7).

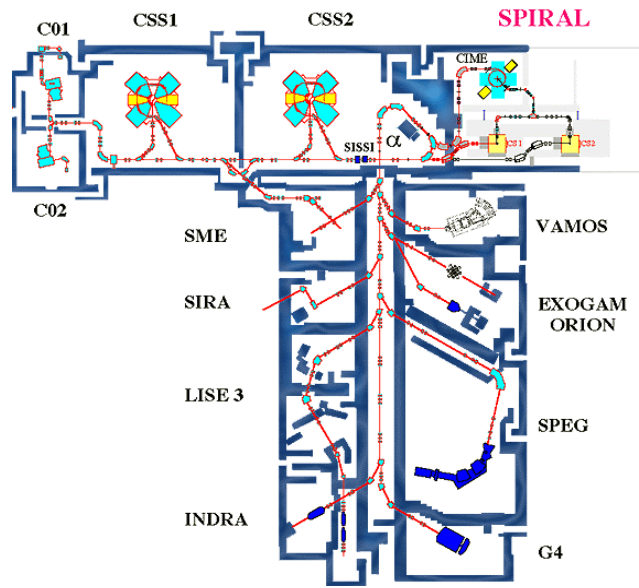

Figure 7: The ISOL facility SPIRAL at GANIL.

At LNS in Catania, Italy, the superconducting K=800 heavy-ion cyclotron is used in the EXCYTE facility to produce ISOL RIBs, post-accelerated in a 15-MV tandem to ~0.2–8 A MeV. The DRIBS project at Dubna generates RIBs with the U400M cyclotron (or via photofission with the M25 microtron), for post-acceleration in the U400 cyclotron.

## INTERMEDIATE FACILITIES

In Europe, a number of projects are proposed or already under way that are expected to make a considerable improvement in the yields of exotic beams available. Amongst these is MAFF in Munich, in which high fluxes of thermal neutrons ($10^{14}$/cm$^2$s) from a reactor will create huge numbers of fission fragments, which can then be post-accelerated in a linac to 7 A MeV.

Another is the SPES project at LNL, Legnaro, in Italy, where an independently-phased superconducting linac will accelerate proton beams to 100 MeV towards the ISOL target. RIBs produced by the converter method with a Be target surrounded by UC$_x$ will be post-accelerated to ~15 A MeV in the existing superconducting linac ALPI.

A third is the SPIRAL 2 project [18], now in an advanced planning stage, in which a linac driver will accelerate either deuterons or light heavy ions onto suitable targets, with the existing CIME cyclotron as post-accelerator (Fig. 8). This can also benefit from the mass-selection capabilities of a cyclotron. One problem which can be anticipated here is servicing of the cyclotron if it becomes too activated by the increased yields of RIBs. It is expected that SPIRAL 2 will yield ~$10^9$ ions/s of $^{132}$Sn.

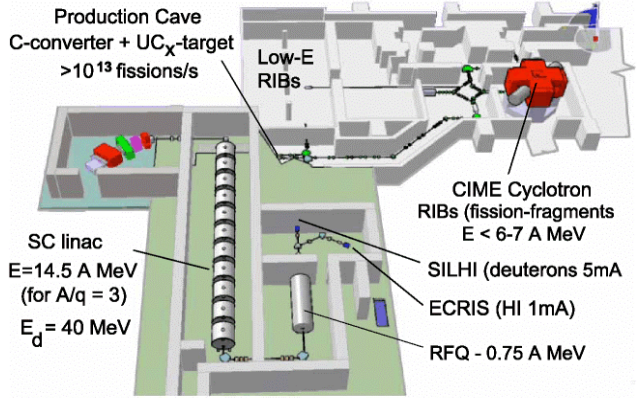

Figure 8: The SPIRAL 2 project at GANIL.

## "NEXT-GENERATION" FACILITIES

### FAIR - Facility for Antiproton & Ion Research

The planned development on the GSI site involves an enormous new facility [19,20] using the existing SIS ring (upgraded) as the injector to two further rings, SIS 100 and SIS 300, respectively (Fig. 9). RIB production will be done using the in-flight technique, and a new high-acceptance fragment separator, the "Super-FRS", is also planned.

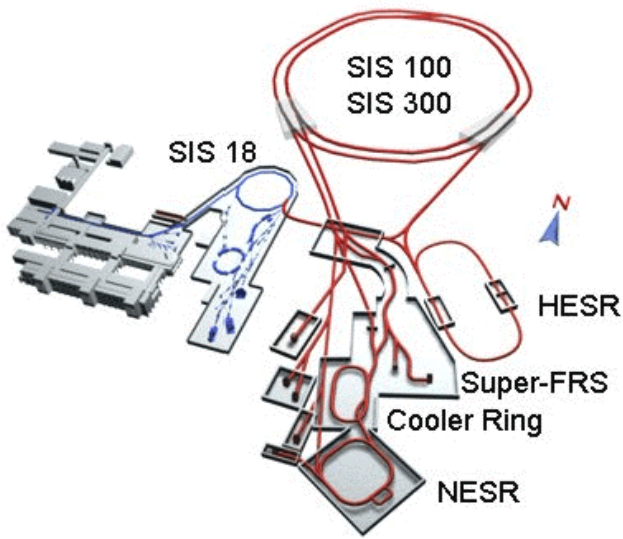

Figure 9: The planned FAIR development at GSI.

With FAIR, uranium ions will be accelerated to 2 GeV/u ($^{238}$U$^{28+}$) at a rate of $10^{12}$/s, a factor of 100-1000 times the present intensity, while $10^9$/s ions of $^{238}$U$^{92+}$ will reach 34 GeV/u. A broad range of radioactive beams is expected, with yields up to factor 10 000 more than at present. A much-improved "Super-FRS" fragment separator is envisaged with a large acceptance needed to utilise as many exotic ions produced as possible (Fig. 10).

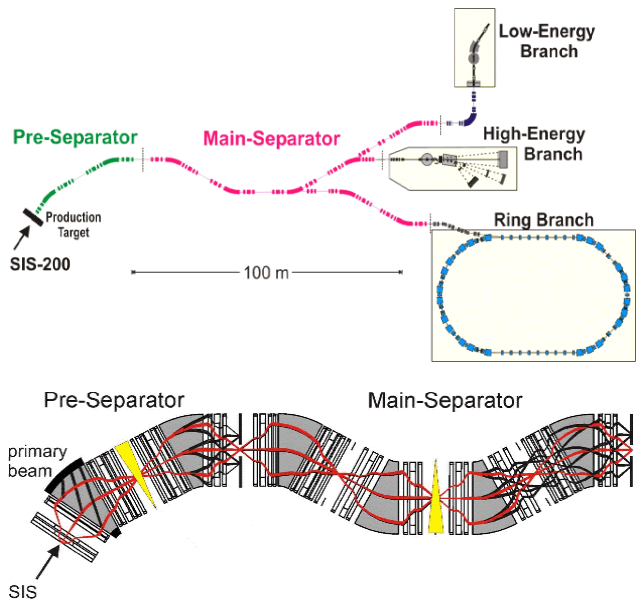

Figure 10: The "Super-FRS" fragment separator layout (top), and the optics of isotope separation (bottom).

### THE EURISOL PROJECT

Within Europe, the FAIR project will use the in-flight technique, and the EURISOL project [21] (Fig. 11) aims to use the complementary ISOL technique to reach other parts of the chart of the nuclides. By using a 1-GeV superconducting proton driver linac capable of producing up to ~5-MW beams, it should be possible to provide yields of exotic RIBs which are orders of magnitude higher than existing facilities.

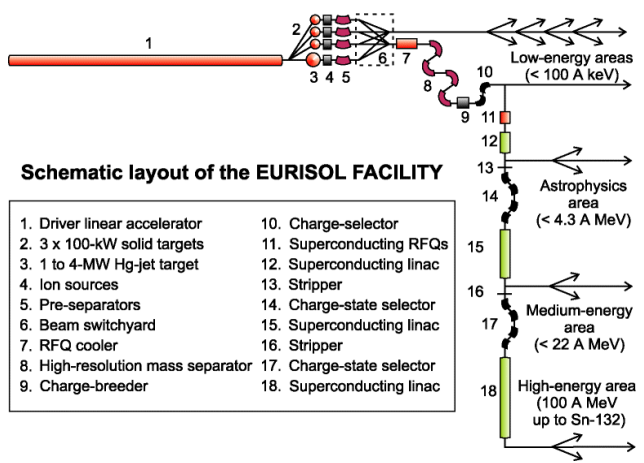

Figure 11: Schematic layout of the EURISOL project.

A number of different target/ion-source areas are envisaged for EURISOL, using (a) 100-kW proton beams on thick solid targets to produce RIBs directly, (b) protons on a "converter" target such as UC$_x$ to release high fluxes of neutrons to produce RIBs in a fission target, and (c) a 1-5-MW beam of protons passing through a mercury-jet target to generate even higher yields of neutrons. Since some of these methods will produce very neutron-rich nuclei, the EURISOL project should be seen as complementary to FAIR. There are of course many assumptions made in yield calculations, and comparisons are therefore difficult. However, for the doubly-magic neutron-rich $^{132}$Sn nucleus, for example, it is estimated that EURISOL can yield up to $10^{13}$ ions per second, compared to around $10^8$/s from FAIR, while for the proton-rich nuclei like $^{100}$Sn nuclei the ISOL method is much less favourable. High yields of very neutron-rich nuclei like $^{132}$Sn, accelerated to energies around 100 A MeV in a superconducting post-accelerator will make it possible to create even more exotic neutron-rich nuclei by secondary fragmentation, and with very acceptable yields.

For EURISOL, the post-accelerator of choice is the superconducting linear accelerator, for reasons of modularity, and comparative ease of steering the beam. However, it should be noted that a preliminary cost estimate [21] showed that a hypothetical superconducting cyclotron capable of accelerating $^{132}$Sn to 100 A MeV could actually be considerably cheaper. Such a machine would also provide mass-selection capabilities, but would exceed the size and mass of any superconducting cyclotron yet built. (The cyclotron option would also have a lower-energy limit, and a small linac would be needed to fill this gap.)

Development of high-power, multi-MW targets will be the most immediate problem, and will draw on the experience of the MEGAPIE group at PSI (Pb-Bi liquid-metal targets) and ongoing research into spallation neutron sources like the SNS and the Japanese Joint Project (liquid-metal targets). A EURISOL Design Study has very recently been approved for funding by the European Commission over the next 4 years, involving some 25 laboratories. A site has not yet been selected for EURISOL, but it would seem logical to locate it at one of the large laboratories in Europe currently involved in RIB developments (e.g. GANIL, LNL Legnaro or CERN).

## BETA BEAMS

A recent idea proposed in Europe is that of generating "beams" of pure electron neutrinos and anti-neutrinos which accompany $\beta^-$ and $\beta^+$-decay of exotic $^6$He and $^{18}$Ne ions respectively, from a RIB factory like EURISOL [22] (Fig. 12). The large accelerating and decay rings needed for this could clearly have important site implications.

I wish to thank all those whose work has contributed material which made this presentation possible.

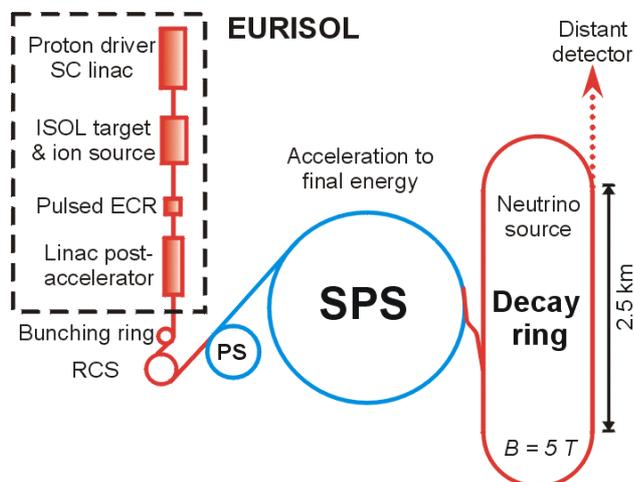

Figure 12: The "beta-beam" concept, illustrated here by coupling EURISOL to the CERN PS and SPL rings.